\documentclass[twocolumn,preprintnumbers,prl]{revtex4-1}
\pdfoutput=1
\usepackage{amsmath,amssymb,graphicx,bm,color}

\begin{document}

\title{CMS Hardware Track Trigger: New Opportunities for Long-Lived Particle Searches at the HL-LHC}

\author{Yuri Gershtein$^a$}

\affiliation{$^a$Department of Physics and Astronomy, Rutgers University, Piscataway, NJ 08854, U.S.A.}

\begin{abstract}
  The planned upgrade of the CMS detector for the High Luminosity LHC allows to find tracks in the silicon tracker 
  for every single LHC collision and use them in the first level (hardware) trigger decision.
  So far, studies by CMS collaboration concentrated on the maintaining the overall trigger performance in the 
  punishing pile up environment. We argue that the potential capabilities of the track trigger are much wider,
  and may offer groundbreaking opportunities for new physics searches. As an example, and to facilitate 
  community discussion, we use a simple toy simulation to study rare Higgs decays into new particles 
  with lifetime of order of a few mm.
\end{abstract}
\maketitle

\section{Introduction}

The CMS detector will undergo extensive upgrades \cite{cmstp} for HL LHC running. A central feature
of the upgrade is a new silicon tracker which allows track reconstruction for every LHC bunch crossing (@40 MHz). 
The first challenge, and the main reason this has never been done before, is to be able to read out the huge 
number of silicon hits within tight latency constraints.
In CMS, thanks to the strong magnetic field, it is possible to construct a tracker that can reliably separate 
small fraction of the hits left by high momentum tracks, and only read out those for track finding at the first level of the trigger (L1).
It is achieved by making tracker modules out of two closely spaced sensors and an integrated circuit 
that correlates the hits in them, providing both coordinate and transverse momentum measurement. The latter assumes
that the track originated at the beam line. The hit pairs in a module are referred to as {\it stubs}.

The $p_T$ selection for stubs to be read out is determined by the bandwidth from the detector 
to the back end electronics, and is fixed at about 2 GeV.  Finalizing the choice of track finding algorithm and hardware
may take a few more years. 
In the meantime, it is imperative to understand
what kind of physics opportunities such track trigger could open up, beside maintaining the overall 
trigger performance at HL LHC environment. 

The goal of this note is to attract community's attention to this topic, and provide an example of a mainstream physics area
that would benefit from extension of the track trigger to off-pointing tracks.
While proper simulation and modeling of the trigger is complicated, a simple toy simulation is sufficient to develop intuition 
and identify areas of interest.

\begin{figure}[b]
\begin{center}
{\bf a)}\\
\includegraphics[width=0.4\textwidth]{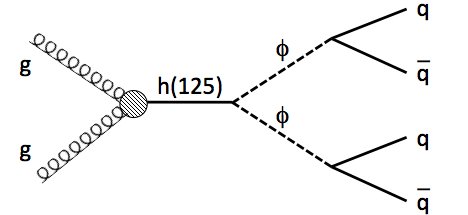}\\
{\bf b)}
\includegraphics[width=0.5\textwidth]{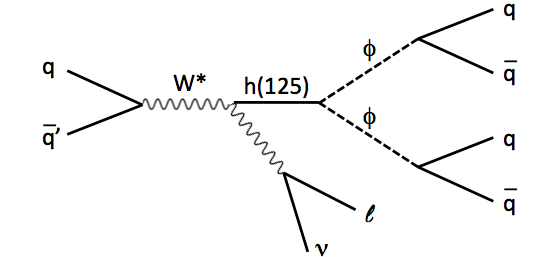}
\end{center}
\caption{\label{fig:hprod} Two final states considered for getting $h\rightarrow\phi\phi$ events on tape: a) gluon fusion production using off-pointing track trigger; b) associated production using single lepton trigger.}
\end{figure}

This note considers all-hadronic final states with low $H_T$, taking SM Higgs decays into four jets (see Fig. \ref{fig:hprod} a) as an example.
Theoretical motivation to look for such decays is very strong, see \cite{exoh} for a comprehensive review. The goal is to probe very small
branching fractions; in this note we assume $Br[h\rightarrow\phi\phi\rightarrow 4q] = 10^{-5}$. 
For prompt decays, the background is overwhelming, but if the $\phi$ has $c\tau$ of a few mm, the offline analysis has very low 
backgrounds \cite{cmsLL}.
The problem is in getting such events on tape, in particular through L1. Below, we estimate how an off-pointing track reconstruction 
at L1 can help. To estimate the efficacy of the approach, we compare it with the best alternatives in absence of off-pointing track trigger: using
associated Higgs production with a W that provides a lepton trigger (Fig. \ref{fig:hprod} b) or considering L1 calorimeter jets with no associated 
prompt tracks. We also speculate on the comparative sensitivity of LHCb experiment to this decay.

\section{Toy Simulation}

\begin{figure}
\begin{center}
\includegraphics[width=0.4\textwidth]{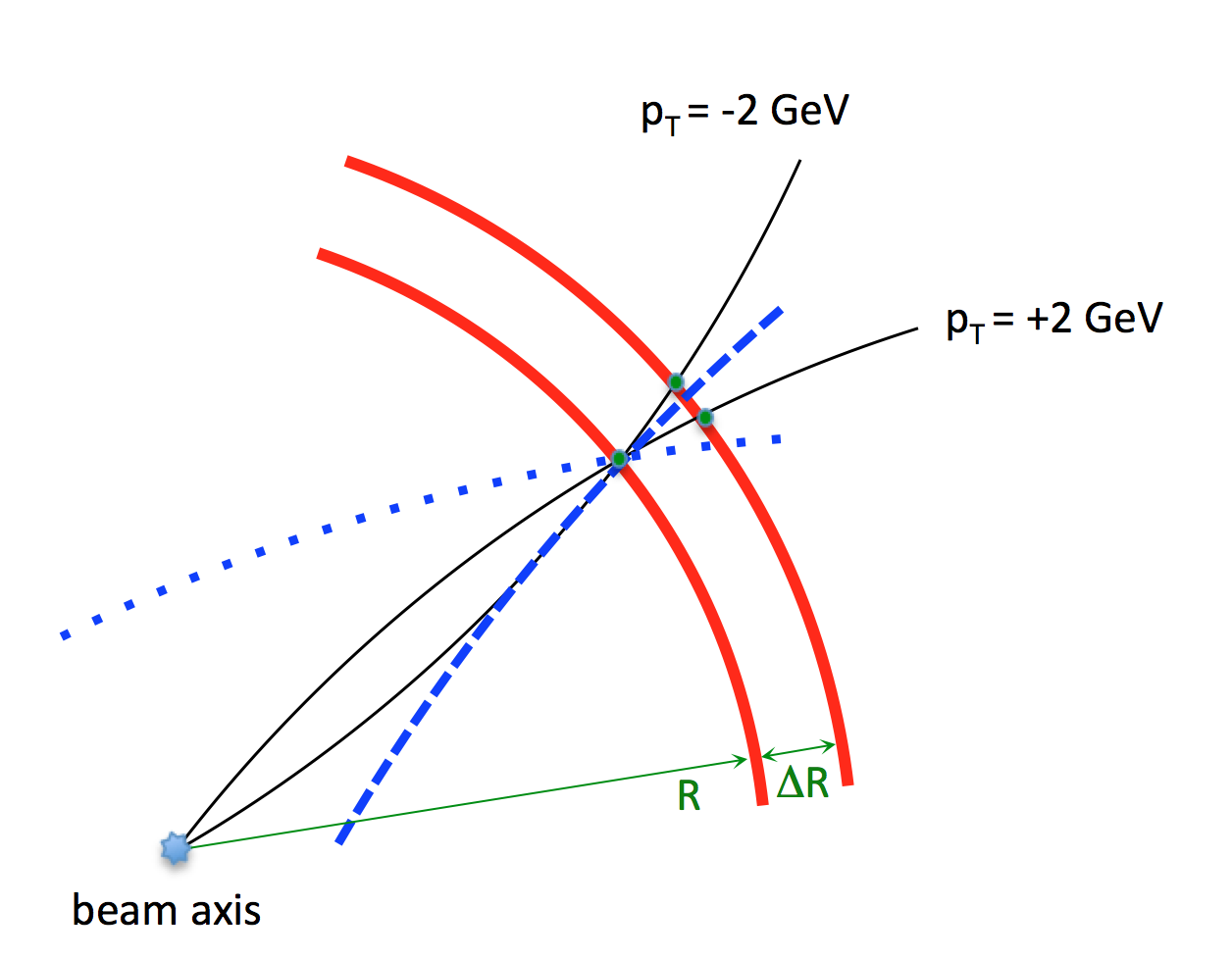}
\end{center}
\caption{\label{fig:geom} Sketch of the toy stub formation in a doublet layer. Four tracks are passing through the same point in the inner layer.
Only the tracks hitting the outer layer between the two green points would produce a L1 stub. The dashed track does, and the dotted does not.
The $R$ and $\Delta R$ values for the six doublet layers in the simulation are 23, 36, 51, 68, 88, and 108 cm and 
 0.26, 0.16, 0.16, 0.18, 0.18, and 0.18 cm.}
\end{figure}

The toy tracker has six perfectly cylindrical double layers \cite{geom} covering $|\eta|<2.4$. 
For each layer, the allowed offset between the two measurements
is below the one expected from 2 GeV prompt tracks. The sketch in Fig. \ref{fig:geom} shows four tracks traversing a double layer: positively and negatively charged 
prompt 2 GeV tracks, and two off-pointing tracks. Dashed track would make a L1 trigger stub, and the dotted one would not. 

\begin{figure}[h]
\begin{center}
\includegraphics[width=0.4\textwidth]{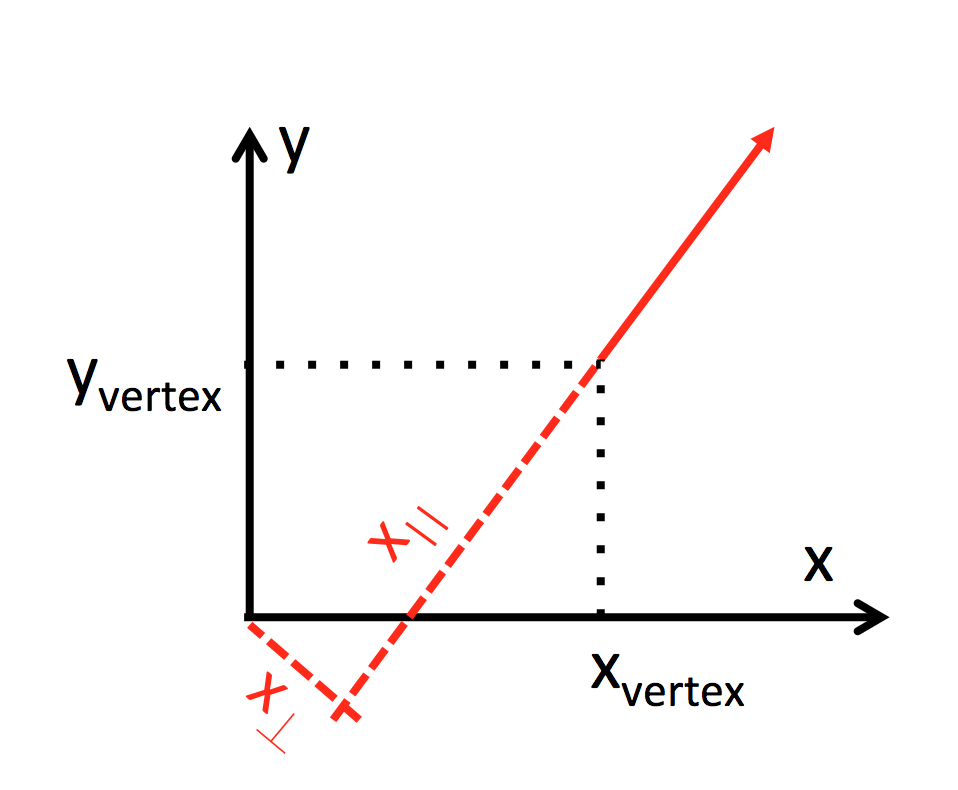}
\end{center}
\caption{\label{fig:defs} Definition of the offsets $x_\parallel$ and $x_\perp$. Solid red arrow is the momentum direction at the 
production point ($x_{\mbox{vertex}},y_{\mbox{vertex}}$).}
\end{figure}

\begin{figure}[h]
\begin{center}
\includegraphics[width=0.5\textwidth]{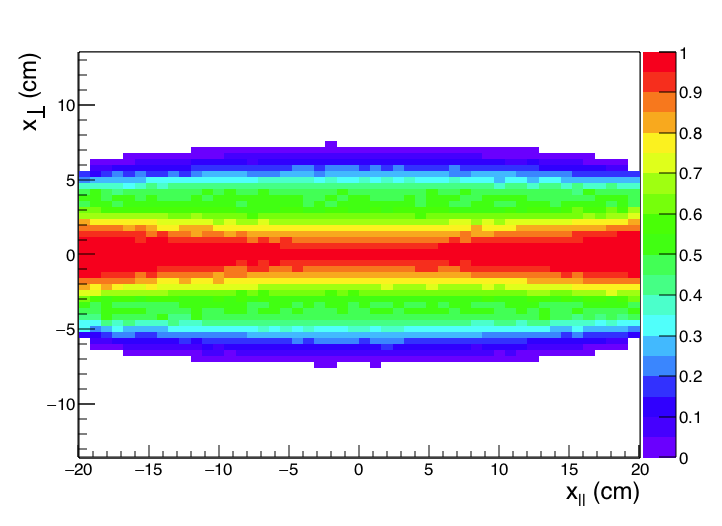}
\includegraphics[width=0.5\textwidth]{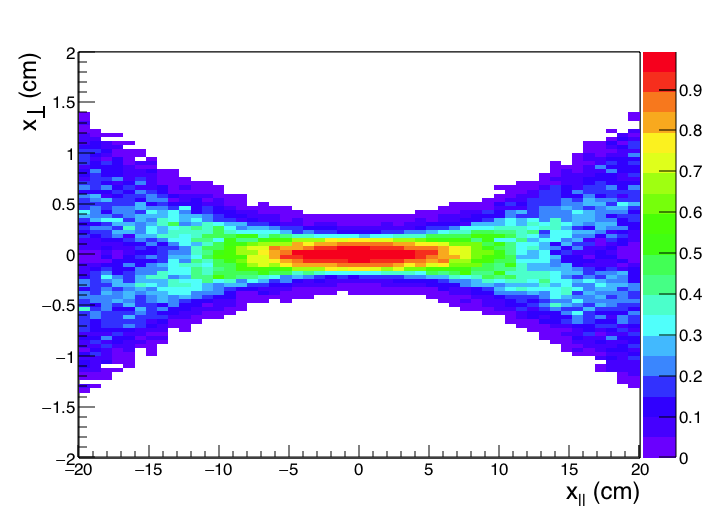}
\end{center}
\caption{\label{fig:trackeff} Efficiency to reconstruct loose (top) and tight (bottom) tracks with 5 or more stubs as a function of the 
uniformly generated particle origin. 
Particle $p_T$ are distributed as expected from low $E_T$ jets, and required to exceed 2 GeV. See text for details. }
\end{figure}

\begin{figure}[h]
\begin{center}
\includegraphics[width=0.5\textwidth]{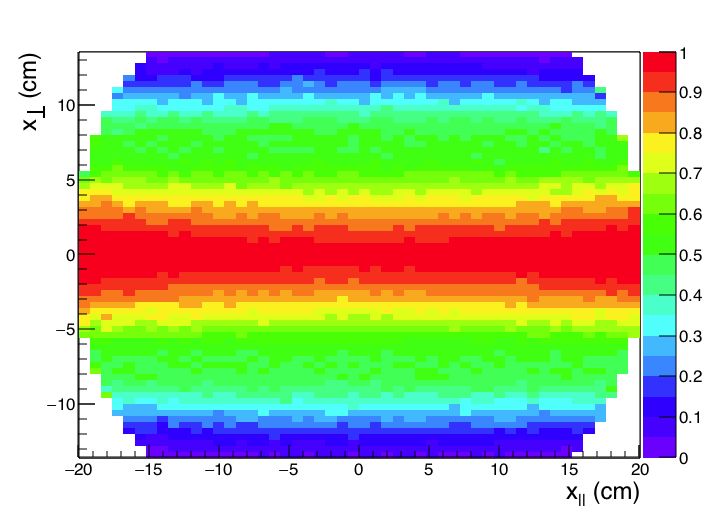}
\includegraphics[width=0.5\textwidth]{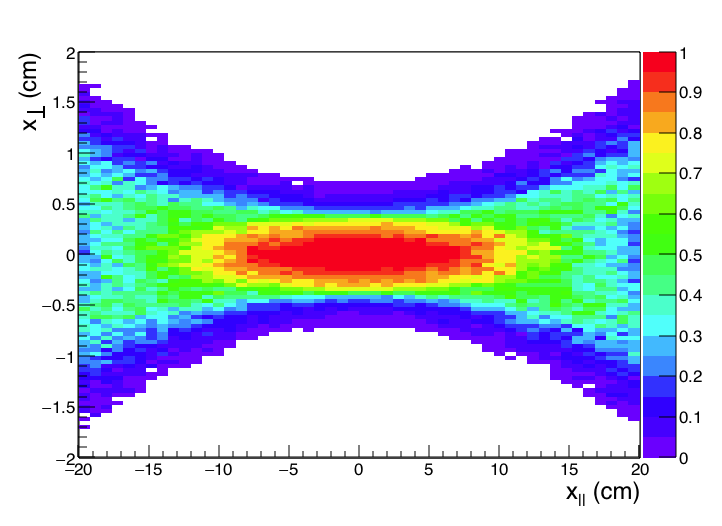}
\end{center}
\caption{\label{fig:trackeff4} Efficiency to reconstruct loose (top) and tight (bottom) tracks with 4 or more stubs as a function of the 
uniformly generated particle origin. 
Particle $p_T$ are distributed as expected from low $E_T$ jets, and required to exceed 2 GeV. See text for details.}
\end{figure}

Two extensions of track finder are considered. {\it Loose} tracks are only required to have a minimal number of stubs. 
{\it Tight} tracks are obtained by fitting the stubs they produced to a circle constrained to the beam line. The number of stubs on a tight track
is the number of stubs deviating from that circular fit by less then 3 strips (300 microns). Tight tracks is a generous approximation for 
an algorithm that assumes prompt production when building a track and allows for non-zero impact parameter for track fit.
For loose tracks, both track building and fitting assumes non-zero impact parameter. We only consider the transverse plane of the track finding
since that's the plane in which the displacement is measured more precisely. We assume that the hits on a track are also linked in the $rz$, but do 
not rely on it for calculation of displacement. 
There is little doubt that such extensions to the track trigger
are technically feasible, but they definitely would be more costly. The discussion here is not on how big the cost increase might be, but on 
whether there is a compelling physics case to pursue it.

One can get an idea for what kind of the range of track displacements produce detectable tracks by looking at 
charged particles above 2 GeV in 40-80 GeV jets (PYTHIA 8 \cite{pythia} is used for generation of all processes in this note).
We uniformly offset the tracks from the beam line to map out the efficiency as a function of a displacement.
The offset is specified along the original track momentum ($x_\parallel$) and perpendicular to it ($x_\perp$), as shown in Fig. \ref{fig:defs}.
Figures \ref{fig:trackeff} and  \ref{fig:trackeff4} show the efficiency to reconstruct loose and tight tracks with $\geq 5$ and $\geq4$ stubs respectively.

Evidently, tight tracks die out beyond $x_\perp$ of a couple of mm, while loose tracks can still be reconstructed at several cm. 
In what follows we consider tracks with at least four stubs, and require some of the tracks to have five or more stubs.

An important aspect that this toy simulation does not address is the rate of fake tracks.
The fake rate was studied \cite{cmstp} and found to be small after the requirements on the quality of the fit,
even for tracks with 4 stubs.
Here, the vertex constraint is removed, presumably leading to the increase in fake rate. However, we are not interested in {\it single} tracks. 
For a track jet the probability to have a fake track is reduced because fake tracks point in random directions and do
not cluster. In what follows, we assume that the fake tracks can be rid of by requirements on the fit quality.

\begin{figure}[h]
\begin{center}
\includegraphics[width=0.5\textwidth]{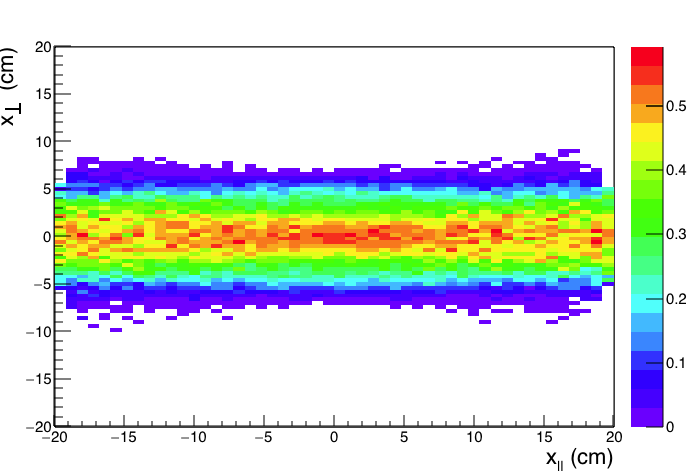}
\includegraphics[width=0.5\textwidth]{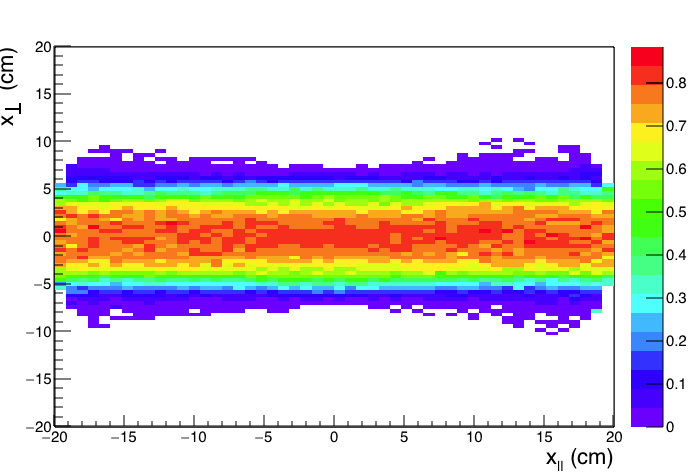}
\end{center}
\caption{\label{fig:jeteff} Efficiency to reconstruct a jet with $p_T$   between 20 to 40 GeV (top) and 40 to 80 GeV (bottom) as a track jet with 3 or more {\bf loose} tracks, 2 of which with 5 or more stubs, and sum $p_T$ of all tracks above 10 GeV, as a function of uniformly generated jet origin. }
\end{figure}

\begin{figure}[h]
\begin{center}
\includegraphics[width=0.5\textwidth]{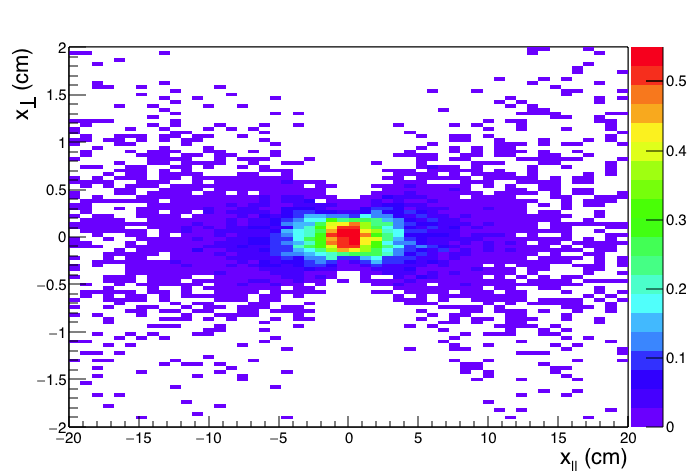}
\includegraphics[width=0.5\textwidth]{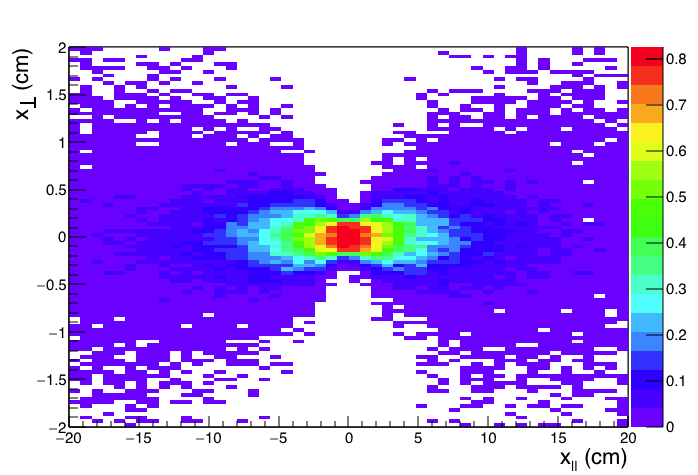}
\end{center}
\caption{\label{fig:jeteff1} Efficiency to reconstruct a jet with $p_T$   between 20 to 40 GeV (top) and 40 to 80 GeV (bottom) as a track jet with 3 or more {\bf tight} tracks, 2 of which with 5 or more stubs, and sum $p_T$ of all tracks above 10 GeV, as a function of uniformly generated jet origin.  }
\end{figure}

Instead of simulating every track in an event, it is more convenient to parametrize a jet response.
An anti-$k_T$ algorithm ($R=0.4$) as implemented in
FASJET \cite{fastjet} is run on the final particles in PYTHIA di-jet events in order to get a generated jet collection.
For each jet, random displacements $x_\parallel$ and $x_\perp$ are generated. 
For each displaced jet, we loop through the charge particle constituents that have $p_T$ above 2 GeV. 
A flat 10\% inefficiency is applied to each particle. The surviving particles are tested against tight and loose track definitions.
Jets are required to contain 3 or more tracks, at least two of which have 5 or more stubs. 
Reconstructed jet $p_T$ is a sum of all track $p_T$'s. 
Figures \ref{fig:jeteff} and  \ref{fig:jeteff1} show the probability for a generated jet to be reconstructed as a jet of loose and tight tracks with $p_T$ above 10 GeV.
This probability is parametrized as a function of $x_\perp$ in bins of of generated jet $p_T$ and $x_\parallel$,
and is used below for the signal yield calculations.

\section{Triggers}

\begin{figure}
\begin{center}
\includegraphics[width=0.5\textwidth]{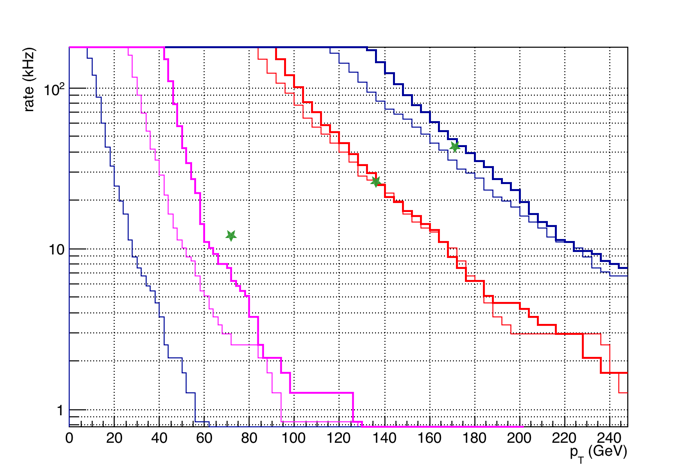}
\end{center}
\caption{\label{fig:trig} Simulated trigger rates of a L1 calorimeter-seeded single jet (blue), di-jet (red) and quad-jet triggers.
Calorimeter jet energy resolution is applied for the thick lines and neglected for the thin lines.
Green stars correspond to results from the  trigger menu in \cite{cmstp}. Quad track jet trigger rate is also shown.}
\end{figure}

The simplified L1 trigger menu in \cite{cmstp} gives expected trigger rates for a single jet, di-jet, and quad-jet triggers.
The jets are found in the calorimeters.
Track trigger is used to determine the jet vertices and the jets are required to originate from the same vertex, greatly reducing
the pile-up effects.
However, starting jet finding with the calorimeter is sub-optimal. It's better to start with the most pile-up resistant system - the tracker.
As an added bonus, triggering on track jets results in events with jets with high charged multiplicity, i.e. easier to vertex.

To come up with reasonable trigger thresholds for track jets, PYTHIA multi-jet events were used.
Figure \ref{fig:trig} shows the obtained trigger rates for single, di- and quad-jet triggers, as well as the quad track jet trigger.
For "calorimeter-seeded" jets, thin lines show spectra for ideal energy resolution. Thick lines assume jet energy smearing
\begin{align}
   \sigma_{p_T} & = \sqrt{N_{PU}^2 + S^2 \cdot p_T} 
\end{align}
with stochastic term $S = 0.9~GeV^{1/2}$ and pile-up noise term $N_{PU}=25~GeV$. 
The cross-section given by PYTHIA was adjusted by 70\% so that the rate of single jet trigger matched number from \cite{cmstp}. 
Given crudeness of our methods, the agreement with \cite{cmstp} is pretty amazing.

\begin{figure}
\begin{center}
\includegraphics[width=0.5\textwidth]{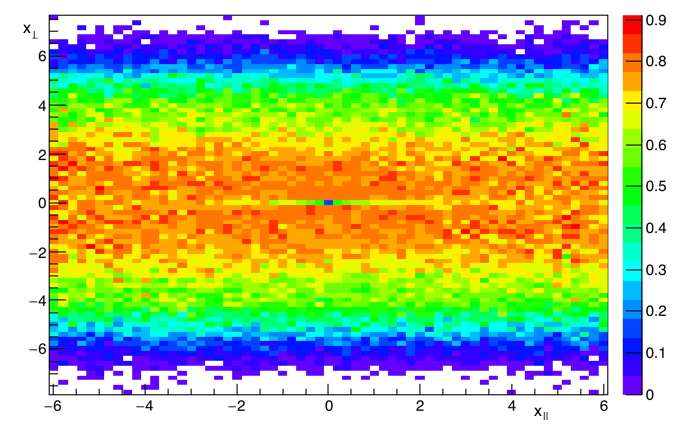}
\end{center}
\caption{\label{fig:jeteff_veto} Same as Fig. \ref{fig:jeteff}b, but with an additional requirement that at least three tracks have 
impact parameters in excess of 300 microns.}
\end{figure}

We conclude that the lowest feasible threshold for a quad track jet trigger is 20 GeV.
However, there is one more handle that can be employed: for displaced jets, tracks should not point exactly at the 
interaction point. The expected impact parameter resolution is about 100 microns. Requiring that at least 3 tracks in a jet
have impact parameters in excess of 300 microns reduces prompt jet efficiency by more then a factor of 50 (see Figure \ref{fig:jeteff_veto}).
It seems therefore that it is not out of the question to run the quad displaced track jet trigger with a threshold of 10 GeV.

There is one more way to tag displaced jets at L1, similar, in spirit, to the ATLAS tags of the decays in the HCAL \cite{atlasEMF}. The latter
result in jets with anomalously low electromagnetic fraction. Requiring a calorimeter jet with no tracks pointing to it
theoretically allows access to shorter lifetimes then that. Unfortunately, while very powerful in absence of pileup, such tracks without jets are
mostly random pileup fluctuations. Nevertheless, one could try being creative with the track vetoes and come up with a 
trigger requiring two such "no-track jets". In this study we look at two thresholds for such trigger, 70 and 100 GeV.
While 70 GeV threshold is undoubtedly very optimistic, this trigger would be an alternative to the track trigger expansion we
argue for, so it makes for a conservative assumption.

To summarize, we consider the following triggers:
\begin {itemize}
\item four jets of loose tracks with $p_T >$ 20 GeV
\item four jets of tight tracks with $p_T >$ 20 GeV
\item four jets of loose displaced tracks with $p_T >$  10 GeV
\item four jets of tight displaced tracks with $p_T>$  10 GeV
\item two no-track jets with $p_T$ above 70 GeV
\item two no-track jets with $p_T$ above 100 GeV
\item single $e$ / $\mu$ triggers with $p_T>$  35 / 18 GeV
\end{itemize}

\begin{figure}
\begin{center}
\includegraphics[width=0.5\textwidth]{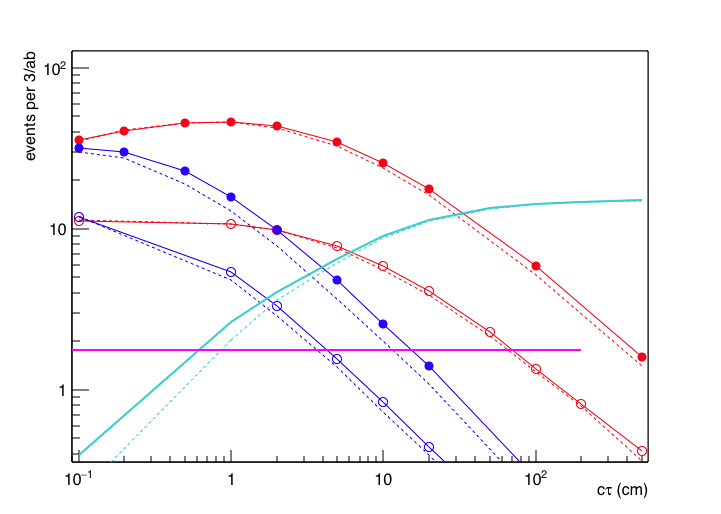}
\end{center}
\vspace{-3mm}
\caption{\label{fig:h125} Event yields for $Br[h\rightarrow\phi\phi\rightarrow 4q]=10^{-5}$ as a function of 
$\phi$ proper lifetime. Red curves correspond to
quad loose track jet trigger, blue - to quad tight track jet trigger. 
Open circles indicate track jets with $p_T$ above 20 GeV, filled circles - displaced track jets above 10 GeV. 
Teal curve corresponds to the  no-track 70 GeV di-jet trigger.
Purple line shows the expected yield from $Wh$ production triggered by a lepton from $W$.
See text for details. }
\end{figure}

\section{Higgs Event Yields}

The $h\rightarrow\phi\phi\rightarrow 4q$ events were generated using PYTHIA. Mass of $\phi$ is taken to be 30 GeV,
and $Br[h\rightarrow\phi\phi\rightarrow 4q]=10^{-5}$. A range of $\phi$ lifetimes is considered, from 1 mm to 5 m.
Proper decay time was randomly generated for each $\phi$ and dilated according to its speed.

We assume that offline analysis selection are similar to the trigger. For $W(\rightarrow\ell\nu)h$ final state, we 
require three of the four jets from $h\rightarrow\phi\phi$ decay to have total charge particle momentum above 10 GeV.
Relaxing the latter requirement to at least two jets from the same $\phi$ increases efficiency by less then 40\% at a
potential cost of non-negligible background contamination.

Figure \ref{fig:h125} shows the expected event yields for different triggers described above.
Jet reconstruction parameters were slightly varied to make sure there are no large variations in efficiency.
For track jets, one (solid lines) or two (dashed lines) tracks with five or more hits were required.
For trackless jets, one track (solid line) or two tracks with total $p_T$ below 10 GeV were allowed to to point along the jet.

While the tight tracks offer substantial increase in sensitivity compared to $Wh$, trigger based on loose tracks  
yields more then a factor of 5 more signal for $c\tau$ of a few mm. No-track jets, even with very optimistic 70 GeV threshold
only become competitive at lifetimes of 50cm or more. 

\section{Event Yields in LHCb}

LHCb experiment \cite{lhcb} will operate at HL-LHC without a hardware trigger and collect 100/fb of integrated luminosity.
While the statistics takes factor of 30 hit, it is still better then the factor of approximately 200 that falling back on 
$Wh$ production at CMS means. A naive PYTHIA based estimate of how many Higgs events with a single $\phi$ decaying in
the LHCb detector fiducial volume, with both daughter quarks $p_T$ above 5 GeV, yields about 15 events, almost an 
order of magnitude better than 1.8 events expected at CMS in $W(\rightarrow\ell\nu)h$ final state.

\section{Other Signals}

\begin{figure}
\begin{center}
\includegraphics[width=0.5\textwidth]{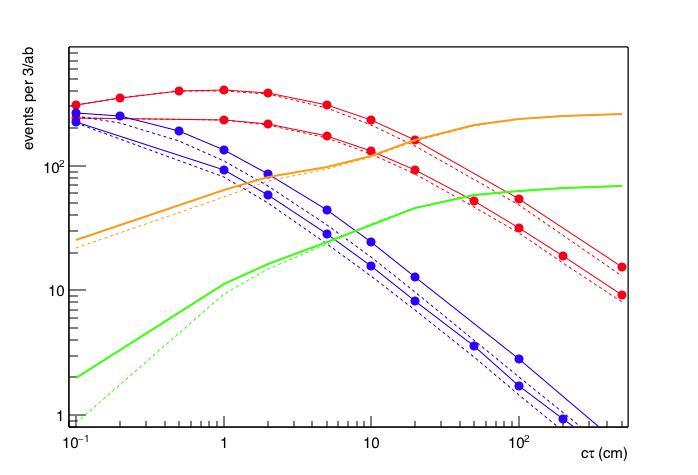}
\end{center}
\caption{\label{fig:h250} Same as Fig. \ref{fig:h125}, but for a $m(H)=250$ GeV and $m(\phi)=60$ GeV.
Number of produced events ($\sigma\cdot Br$) is assumed to be the same as for 125 GeV Higgs.
Brown and green curves are for no-track di-jets above 70 and 100 GeV respectively.}
\end{figure}

The rare SM Higgs decays considered above is truly challenging, because of the small total $H_T$ in the event. 
The efficacy of proposed triggers increases very quickly with mass of the objects produced.
As an example, we doubled the masses of the particles in the last section to $m(H)=250$ GeV and $m(\phi)=60$ GeV.
This is still inaccessible with $H_T$ trigger, whose threshold is not expected to be below 350 GeV at L1 \cite{cmstp}.
The trigger efficiency rises by almost an order of magnitude, with several hundred events expected (Fig. \ref{fig:h250}).

\begin{figure}
\begin{center}
\includegraphics[width=0.25\textwidth]{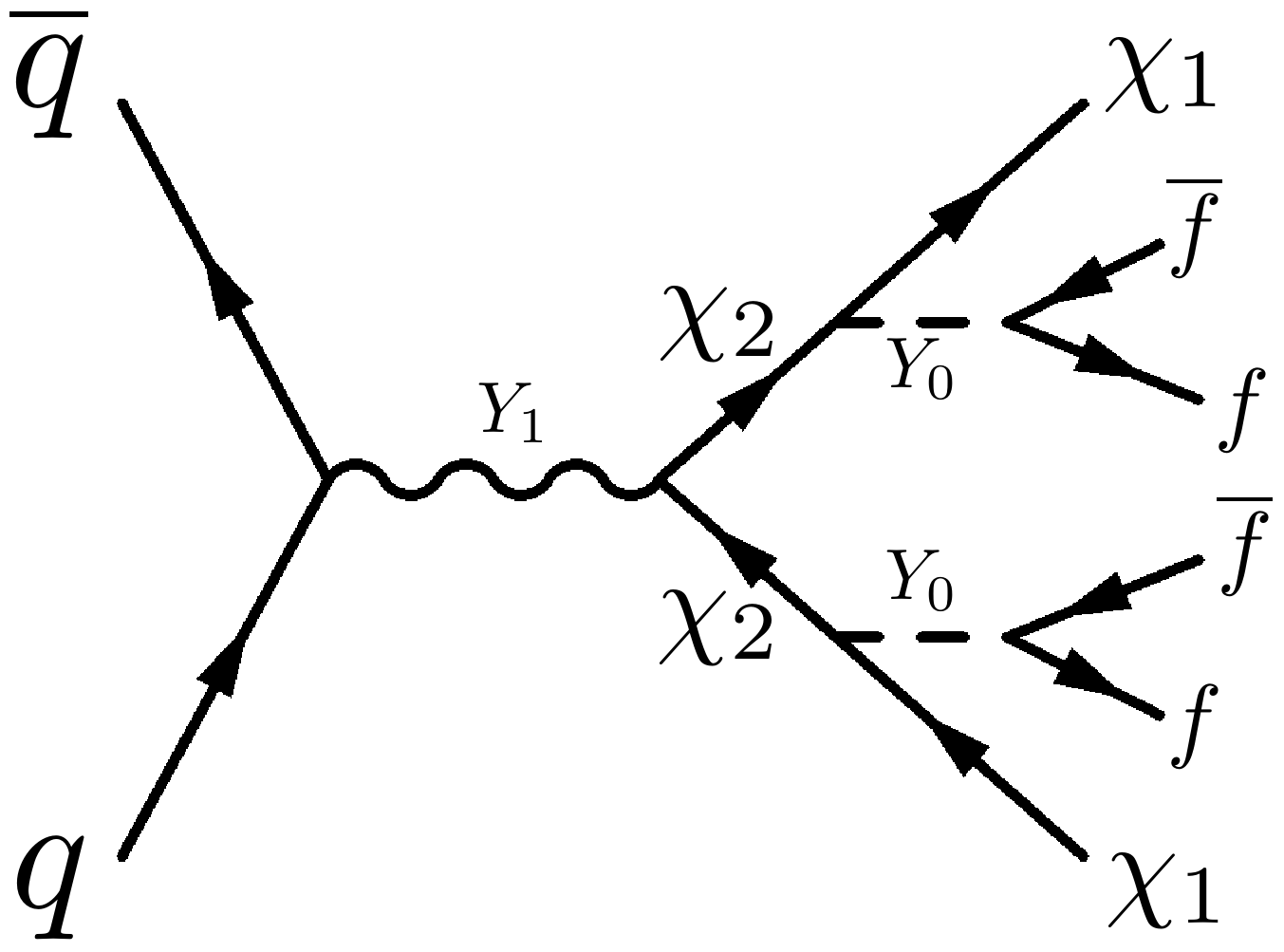}
\end{center}
\caption{\label{fig:dm} An example process from \cite{dm}.}
\end{figure}

While Higgs-like signals may be accessible in associated production, some new physics signals may not.
A good example is a simplified long-lived Dark Matter scenario proposed in \cite{dm}. While that paper focuses on
mass hierarchy like $m(\chi_2)\sim $TeV, $m(\chi_1)\sim $GeV, and $m(Y_0)\sim 100$GeV, it is just as plausible that 
$\chi_2$ and $\chi_1$ are both heavy with small mass splitting so that the two $Y_0$ are not ultra-relativistic and 
result in jetty events with small $H_T$ and missing $E_T$. Those will benefit a lot from the displaced track trigger.

\section{Summary}

CMS experiment has heavily invested in a silicon tracker that allows reconstruction of high $p_T$ tracks at L1.
While the primary stated purpose of track finding at L1 is to preserve the trigger performance in HL LHC environment,
an ability to find tracks at L1 can provide a new lamppost for new physics searches.
In this note, we argue that technically feasible extensions of track trigger could provide large increases in 
sensitivity to long-lived particles, in particular in exotic Higgs decays.

\begin{acknowledgments}
The author is grateful to Scott Thomas, Marco Farina, Eva Halkiadakis, Kristian Hahn, and Michael Williams for valuable discussions and feedback.
The work is partly supported by the NSF grant number 1607096.
\end{acknowledgments}


\begin{thebibliography}{99}
\bibitem{geom}
Inner radii of the doublet layers are 23, 36, 51, 68, 88, and 108 cm. 
The spacing between inner and outer layers in the doublet layers 
are 0.26, 0.16, 0.16, 0.18, 0.18, and 0.18 cm respectively.
This approximates the barrel geometry in \cite{cmstp}.
\end{thebibliography}
\end{document}